\documentclass[a4paper,12pt]{article}
\usepackage{geometry}
\usepackage{caption}
\usepackage{amsmath}
\usepackage{amssymb}
\usepackage{graphicx}
\usepackage{bm}
\usepackage{mathptmx}
\usepackage{etoolbox}
\usepackage{relsize}
\usepackage{mathtools}
\usepackage{xcolor}
\usepackage{todonotes}
\usepackage{url}
\usepackage[affil-it]{authblk}
\usepackage{hyperref}

\begin{document}

\title{Transformation Cost Spectrum for \\ Irregularly Sampled Time Series}
\date{}

\author[*]{Celik Ozdes}
\author[*]{Deniz Eroglu}

\affil[*]{\small Faculty of Engineering and Natural Sciences, Kadir Has University, Istanbul}

\maketitle

\abstract{
\begin{center}
\begin{minipage}{0.75\textwidth}
\small
Irregularly sampled time series analysis is a common problem in various disciplines. Since conventional methods are not directly applicable to irregularly sampled time series, a common interpolation approach is used; however, this causes data distortion and consequently biases further analyses. We propose a method that yields a regularly sampled time series spectrum of costs with minimum information loss. Each time series in this spectrum is a stationary series and acts as a difference filter. The transformation costs approach derives the differences between consecutive and arbitrarily sized segments. After obtaining regular sampling, recurrence plot analysis is performed to distinguish regime transitions. The approach is applied to a prototypical model to validate its performance and to different palaeoclimate proxy data sets located around Africa to identify critical climate transition periods during the last 5 million years and their characteristic properties.
\end{minipage}
\end{center}
}
\vspace{15mm}
\normalsize

\section{Introduction}

Time series analysis techniques have been employed for decades to extract the characteristics of the data sets to tackle various challenges and explain phenomena occurring in natural, social, and engineering sciences \cite{anderson1977box}. This progress significantly helps us understand past variations in associated systems and the underlying mechanisms behind them, allowing us to make goal-directed predictions \cite{friedman2001}.\\

Traditional time series analysis methods are designed to perform on regularly sampled data, i.e., the time resolution $\Delta t = t(s_i+1)-t(s_i) =$ const $\forall i \in [0,N-1]$ \cite{livina2007,trulla96,small2013complex}. However, data sets are naturally collected with irregular spacing in several disciplines such as astrophysics and earth sciences due to data point scanning costs or lack of high-quality observation. For example, in palaeoclimate science, time series are retrieved from measuring historical archives such as stalagmites, tree rings, and lake sediments \cite{marwan2021,eroglu2014a,rehfeld2013,eroglu2016}.The measurements on sediment proxies are performed on an equidistant length axis but due to the changing growing rate of structures (sedimentation rate), the corresponding inferred temporal axis is not equidistant. Thus, these measurements result in irregularly sampled time series. Moreover, scanning data points is a prolonged and expensive process from such data sources and it is preferable to scan only the most reasonable time spans and sometimes selectively for the research aims and this may result in non stationary sampling rates.\\

Interpolation is a common approach to preprocess such irregularly sampled time series to analyze it with traditional methods. However, interpolation techniques generally cause harmful data reconstruction by replacing all data points with a given arbitrary function \cite{rehfeld2011,rehfeld2014}. The palaeoclimate data sets naturally have poor quality and need a preprocessing method to regularize the sampling while protecting the data's sheltered information of natural climate events.\\

The TrAnsformation Cost Time Series (TACTS) method was recently introduced to regularize time sampling of data with low bias \cite{ozken2015,ozken2018}. Although TACTS is a promising data preprocessing method, it reduces the number of time points by converting consecutive data segments using a metric distance. Palaeoclimate proxies are generally sparse data sets; reducing the total number of points is not preferable for the analysis. Furthermore, TACTS has an arbitrary segment size, $\omega$, to compute the distance between consecutive segments. There is no unique way to set this segment size, and varying this parameter causes changes in the resulting cost time series. \\

This current study introduces a modified TACTS method to produce a spectrum of cost time series in the desired length. Note that the spectrum construction eliminates the heavy dependency on the arbitrary segment size, $\omega$. The manuscript is organized as follows: in the following section, we introduce the modified TACTS technique to construct a spectrum of cost time series and highlight how the parameters of the cost-transformation function can be determined based on measurement data. In Sect.~\ref{sec:ts}, we discuss how to analyze the produced cost spectrum time series using recurrence plot \cite{marwan2007} and recurrence quantification measures. In Sect.~\ref{sec:apps}, we apply our method using numerical data from a paradigmatic model system, the logistic map, with irregular sampling and measurement noise. We then analyze paleoclimate records surrounding Africa over the past 5,000,000 years as a real-world application. Finally, we express a conclusion in Sect.~\ref{sec:conclusion}.

\section{Data preprocessing in time series analysis}

TACTS method was introduced to regularize time series with minimum bias~\cite{ozken2015,ozken2018}, which was motivated from the following Refs.~\cite{victor1997,suzuki2010}. Here, we present the TACTS technique along with a spectrum analysis to detect different types of bifurcations and distinguish various periodic windows.

\subsection{Transformation cost time series}
In TACTS, data segments are mapped to points of a metric space where we can define the scalar cost of transformation ($c$) between consecutive segments as the metric. Let $t(\alpha),X(\alpha)$ denote irregularly spaced measurements for a set of events $\{\alpha\}$ in a segment. TACTS is a difference operator on the series of consecutive segments coming from this set and employs the cost of transforming two adjacent segments using a sequence of operations of three kinds: (1) scaling or changing the amplitude of a data point, (2) shifting a data point in time, and (3) ignoring (creating or deleting) a data point. The transformation cost of converting a segment $S_a$ into a non-empty and non-overlapping consecutive segment $S_b$ is given by the infimum of all possible transformation costs. The equations for segment-to-segment and point-to-point transformation costs $c$ and $d$ are as given below:
\begin{subequations}
\begin{align}
c(S_a,S_b) & = \frac{1}{\| I\|}\inf\left\{\lambda \| I - J\|+\mathlarger{\sum}_{\alpha,\beta\in J} d(\alpha,\beta)
\right\}\\
d(\alpha,\beta) & = \lambda_\tau|t_a(\alpha)-t_b(\beta)|+\lambda_x|X(\alpha)-X(\beta)|
\end{align}
\end{subequations}
\noindent where $I = S_a\cup S_b$ is the set of events and $J\subseteq I$ is the set of those events which are shifted and scaled between segments. $\alpha\in S_a \cap J$ and $\beta \in S_b\cap J$ denote two such shifted/scaled events in adjacent segments. $t_a(\alpha)$ and $t_b(\beta)$ are the relative time signatures of points $\alpha$ and $\beta$ in sets $S_a$ and $S_b$ offset by the minimum time of intervals which span these segments. $|.|$ denotes absolute value, and $\|.\|$ denotes the cardinality of a set. Parameters $\lambda_x$ and $\lambda_\tau$ are the cost factors of amplitude and time transformations, and $\lambda$ is the cost of ignoring a point operation. \\

In previous iterations of TACTS method, amplitudes were observed to be strongly correlated with the data sampling rate and a non-stationary sampling rate distribution produced biased results. Braun et al. \cite{braun2021sampling} offer a treatment of these biases using sampling rate constrained (SRC) surrogates. For modified TACTS, we instead introduce a normalization procedure for $c(S_a,S_b)$ to be robust against variable sampling rates by dividing the segment transformation cost by the total number of measurement points in interval $S_a\cup S_b$, in other words, the number of point-wise transformations necessary to match the segments. This modification formulated in Eq.1a is more advantageous because it is preferable to analytically remove the sampling rate dependency so that TACTS remains deterministic. We can consequently use multiple TACTS series with arbitrary resolutions and offsets to tackle the problem of segment size dependency. \\

Now, for a set of evenly sampled time points $T = \{t_0+i\delta ~|~ i\in\mathbb{N}\}$ where $t_0$ is initial time point and $\delta$ is an arbitrary sampling rate, we can evaluate TACTS: $C(t)=c(S_a(t),S_b(t))$ where $S_a(t)$ is the segment of points in $(t-\omega,t)$ and $S_b(t)$ is the adjacent segment containing the points in $(t,t+\omega)$ for some time window size $\omega$. When we set $\omega>\delta$, the segments $S_a$ and $S_b$ overlap and TACTS provides higher resolution output as long as $S_a(t_i))\cap S_a(t_{i+1})\neq \text{\O}$ and $S_b(t_i))\cap S_b(t_{i+1})\neq \text{\O}$, that is, set of points considered in consecutive time points are distinct. Let $T^* = \{t\in T ~|~ S_a(t)\neq\text{\O},S_b(t)\neq\text{\O}\}$ be the subset of $T$ with no empty segments. Also let $D=\{t_{i+1}-t_i ~|~ i\in\mathbb{N}\}$ be the set of time differences for $t(\alpha)$ and $D^* = \{d\in D~|~d<\omega\}$ be those differences which are smaller than a single segment's size. Parameters $\lambda_x$, $\lambda_\tau$ are normalization parameters chosen to be functions of timeline $T$, segment size $\omega$ and data characteristics $X(\alpha)$ and $t(\alpha)$:
\begin{subequations}
\begin{align}
\label{eq:lambda0}
\lambda_x^{-1}&= \frac{1}{|T^*|}\mathlarger{\sum}_{t\in T^*}\left|\frac{1}{|S_a(t)|}\sum_{\alpha\in S_a(t)}{X(\alpha)}-\frac{1}{|S_b(t)|}\sum_{\beta\in S_b(t)}{X(\beta)}\right|\\
\label{eq:lambdat}
\lambda_\tau^{-1}&= \frac{1}{\|D^*\|}\mathlarger{\sum}_{d\in D^*}d
\end{align}
\end{subequations}
\noindent where $\lambda_x$ is the reciprocal of the expected mean amplitude difference between two consecutive non-empty $\omega$-segments, and $\lambda_\tau$ is the reciprocal of the expected time difference between two consecutive time points which are not at the boundary of empty segments (gaps) for a given $\omega$ along timeline $T$. Consecutive gaps result in a transformation cost of zero for consecutive time points in the TACTS series, so we remove these segments from further analyses to eliminate artificially high recurrence in these regions. \\

\begin{figure}[p]
\begin{center}
\includegraphics[width=0.85\textwidth]{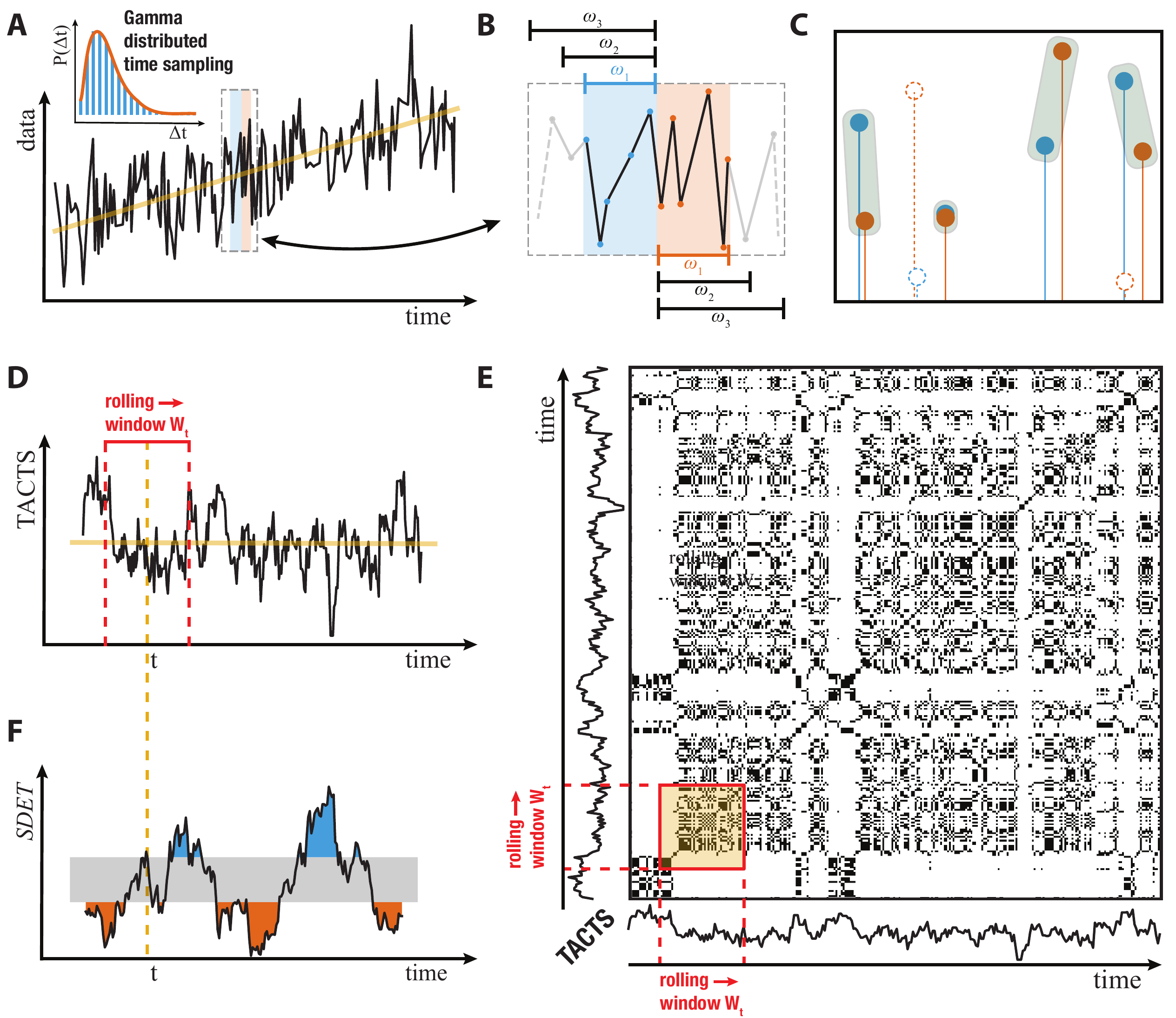}
\captionsetup{width=0.85\textwidth}
\caption{Illustration of the methodology. (A) An irregular time series where the distance between time points has a non-uniform distribution. (B) Segments of length $\omega$ are extracted from the time series and considered two adjacent segments denoted $S_a$ and $S_b$ shown with blue and orange colors. (C) The minimum cost transformation between these segments among all possible transformations between segments. Finding the minimum includes point-wise transformation costs for aligning respective points in time/space dimensions for the points mapped to each other and the cost of ignoring any other point in $S_a \cup S_b$. (D) Generating a spectrum of regular and stationary TACTS time series from these costs corresponding to each choice of $\omega$-window. The TACTS is analyzed using rolling windows $W_t$ with fixed length $|W_t|=L$, suitably chosen to contain enough TACTS points to distinguish different dynamical behavior. (E) A recurrence plot is generated for all TACTS series. Rolling windows analysis gives a single determinism value associated with the time point at the center of our rolling windows. (F) After obtaining an ensemble of DET series from each TACTS, it is possible to use them as more or less independent response signals associated with a particular period or aggregate these series at each time point to a single measure SDET, called a spectrum determinism. Gray horizontal band in (F) is the confidence interval and blue(orange) shadings represent regions with significantly high(low) determinism values.}
\label{fig:method}
\end{center}
\end{figure}

Finally, the cost of ignoring a point, $\lambda$, is the most critical parameter for optimization. Larger $\lambda$ promotes larger shifting and scaling magnitudes for the point-to-point transformations, and smaller $\lambda$ promotes high sampling rate adaptivity for the almost regular sampling. Since modified TACTS is normalized with respect to the number of considered data points and the highest possible cost corresponds to the segment transformation that chooses to ignore all points in the segments, we have $C(t)\le\lambda$. Furthermore, we select $\lambda(\omega,T)$ to minimize the Kolmogorov-Smirnoff (KS) distance of the resulting TACTS series to achieve the most gaussian distribution of the costs $C(t)$. The transformation of an irregularly sampled time series to a cost time series is illustrated in Fig.~\ref{fig:method}(A-D).

\subsection{TACTS spectrum}
In modified TACTS, we consider a set of segment sizes $\{\omega_1,..,\omega_k\}$ and optimize a discrete spectrum of TACTS functions $C_i(t)$, where $i = 1,..,k$. These functions represent the speed of change for their respective embedding of the original data. Each $C_i(t)$ is sensitive to $\omega_i$-periodic patterns in the original time series since the segment states of the interval $\omega_i$ evolve the slowest in $\omega_i$-periodic regimes. The ignoring cost $\lambda_i$ for each $C_i$ is optimized to interpret the particular data sampling concerning different time-frames. As the point sets for segments is also a function of $\omega_i$, unit costs $\lambda_{x,i}$ and $\lambda_{\tau,i}$ are also distinct for each $C_i$. Finally, Each $C_i$ is evaluated along the same - preferably regular - timeline $T$, so we can use quantitative time series methods on each to obtain an aligned analysis for the entire discrete spectrum. While each $C_i$ can also be used individually for different time series analyses, we advocate using ensemble methods on the entire spectrum since we observe that no single segment size $\omega_i$ successfully captures the target dynamics at every point.

\section{Time series analysis}
\label{sec:ts}
After time sampling regularization, the time series can be analyzed by traditional techniques such as wavelets and spectral analysis \cite{priestley1996wavelets}, Lyapunov exponents \cite{kantz1994robust}, entropy-based measures \cite{pincus1991approximate} and visibility graph \cite{lacasa2008time}. Among many data analysis tools, we opt to use recurrence plot methods since we aim to detect regime changes and RP is one of the most used and successful tools for this goal \cite{marwan2007}.

\subsection{Recurrence Plot}
The recurrence plot (RP) is a powerful tool for visualizing dynamical systems' recurrences in their phase space \cite{eckmann1987} and was used to quantify the behavior of dynamical systems in various fields, e.g., earth science \cite{eroglu2016,hirata2010identifying}, chemistry \cite{eroglu2014b}, economics \cite{hirata2012timing}, heart rate variability \cite{zimatore2021recurrence}, traffic congestion \cite{wu2022recurrence}, music studies \cite{proksch2022coordination}, psychology \cite{baranowski2022being}. This current study aims to detect critical transitions in prototypical dynamical systems and paleoclimate records by RP after preprocessing by TACTS.\\

The RP is a matrix containing the Poincar\'e recurrences of phase-space states \cite{poincare1890}. The Poincar\'e recurrence theorem says that a trajectory $\vec{x}_i \in \mathbb{R}^m$ for $i=1, \dots, N$, on sufficiently long and finite time, will revisit the $\epsilon$-neighborhood of a previous state. The RP is a time vs. time matrix of the recurrences and is defined as
\begin{eqnarray}
R_{i,j}(\epsilon) = \Theta(\epsilon - || \vec{x}_i - \vec{x}_j ||) &
i,j = 1, \dots, N
\label{eq:recurrence}
\end{eqnarray}
where $\epsilon$ is some threshold distance, $|| \cdot ||$ is some distance measure, and $\Theta(y)=1$ if $y \geq 0$ and $0$ otherwise \cite{marwan2007}. Although there are some suggested techniques to choose an optimal recurrence threshold $\epsilon$, selecting $\epsilon$  depends on the associated research question.\\

Quantifying recurrence patterns of the RP infers the properties of dynamical systems, and this quantification can capture the regime changes such as extreme events in climate systems \cite{marwan2021}, and economic crises \cite{addo2013}. Among different kinds of structures, diagonal lines of the RP, parallel to the main diagonal (bottom left to top right in Fig.~\ref{fig:method}(E)), indicate the joint period when a trajectory accompanies locally neighboring paths. Therefore, comparing the abundance of the recurrence points forming diagonal lines with the single points in the RP measures the predictability of the dynamical systems. The determinism of RP, $DET$, measures the predictability with the fraction of points that lie in diagonal lines with respect to all points and is given by
\begin{eqnarray}
DET = \frac{\Sigma_{l=l_{min}}^{N}lP(\epsilon,l)}{\Sigma_{l=1}^{N}lP(\epsilon,l)}
\label{eq:determinism}
\end{eqnarray}
where $\epsilon$ is the selected RP threshold and  $P(\epsilon,l)$ is the histogram of diagonal lengths. In this study we use $l_{min} =2$ and RP is computed using the Euclidean distance norm, i.e $\|x,y\|=|x-y|$ for one-dimensional time series, and $\epsilon$-threshold distance is chosen as a fraction of the standard deviation of our TACTS time series, $\epsilon = 0.1\sigma(x)$, to allow for enough recurrences populating the RP. \\

We use sliding windows analysis on our TACTS series by fixing a recurrence frame size $L$ and a regular recurrence timeline $T_R$ and calculating the determinism series $DET(t)$ where $t\in T_R$. $DET(t)$ is the determinism evaluated at time $t$ and comes from the partial recurrence plot associated with data segment $W_t=\{x_i | t_i\in(t-L/2,t+L/2)\}$ (Fig.~\ref{fig:method}(D-F)). For the typical case, $W_t$ are overlapping data segments because recurrence frame $L$ is chosen to be much larger than $\Delta T_R$, timestep for the regular recurrence timeline $T_R$. \\

\subsection{Determinism spectrum}
Treating each series in the spectrum independently, we can measure the determinism series for the TACTS spectrum $C_i(t)$ and use their aggregate determinism values at each time point. Spectrum determinism, $SDET$, is the average determinism of segments from TACTS series with different $\omega_i$ evaluated along some recurrence timeline $T_R$ (see Fig.~\ref{fig:method}(B)),\\ 
\begin{equation}
SDET(t) = \frac{1}{k}\sum_{i=1}^k{DET_i(t)}
\end{equation}
where $DET_i(t)$ are the time series of determinism values corresponding to different TACTS series with distinct $\omega_i$, evaluated at the centers of rolling windows for some fixed recurrence frame size $|W|=L$ (Fig.~\ref{fig:method}(E)).\\

\section{Applications}
\label{sec:apps}
The primary focus of our study is to identify regime changes in a dynamical system that goes through critical transitions. This section shows a prototypical application on the logistic map and a real-world application on a set of palaeoclimate time series around Africa.

\subsection{Logistic Map}
We consider the modified logistic map given by $x_{n+1}=r_n\cdot x_n\cdot (1-x_n)$ to be able to apply a rolling window analysis and get closer to real-world application. We assume that the system parameter $r_n\in[3.5,4]$ slowly drifts as we generate $N=20,000$ time points to simulate a system that changes behavior along its single trajectory. We then remove a fraction of random points from the series $\{x_0,..,x_N\}$ to mimic temporal irregularity. Finally, to portray a noisy measurement process as encountered in real data, we add constant uniform noise, $|\xi_i| \le K\sigma(x)$ to the measurements bounded by a fraction $K$ of the standard deviation of time series data.\\

\begin{figure}[h!]
\begin{center}
\includegraphics[width=0.9\columnwidth]{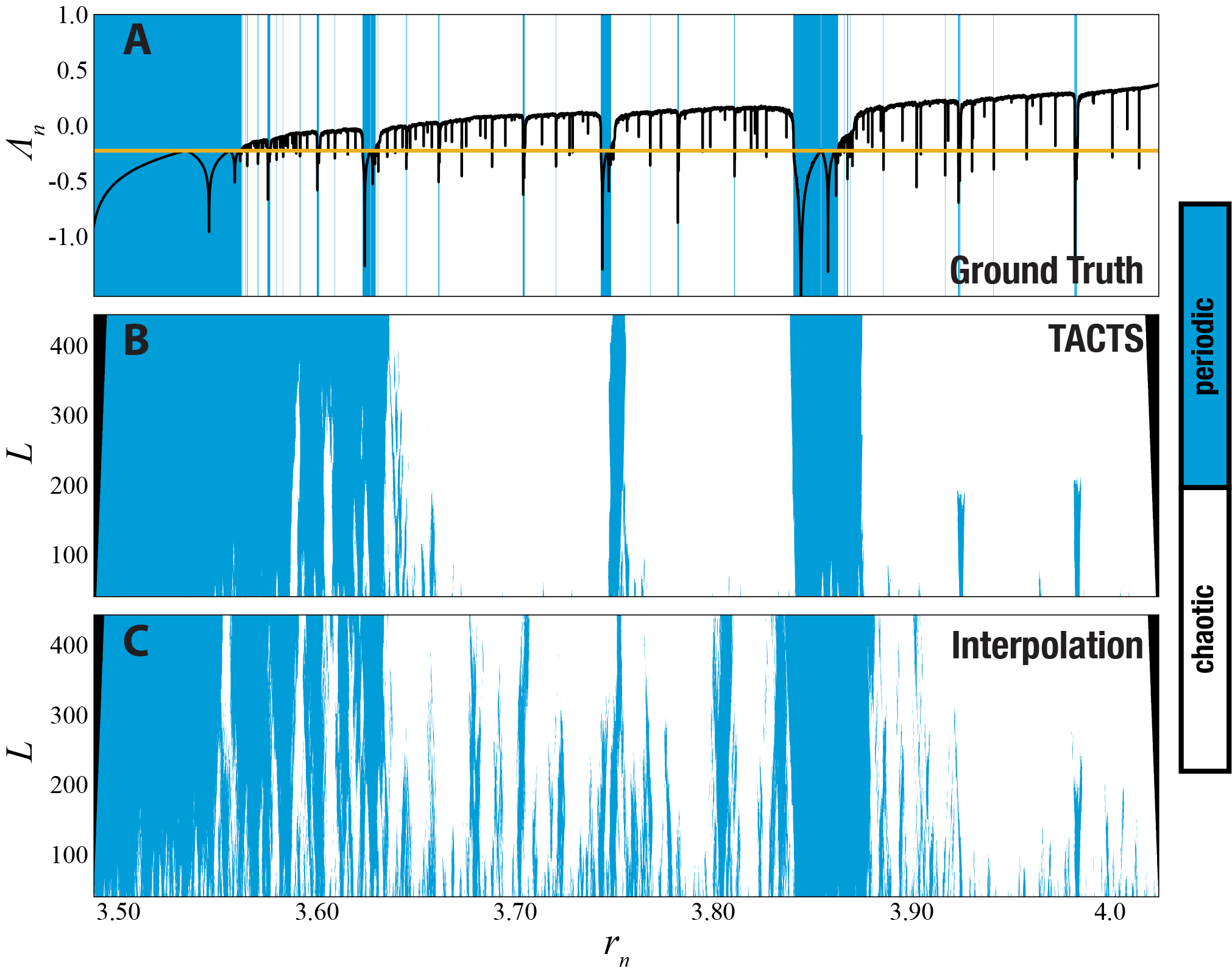}
\captionsetup{width=0.9\textwidth}
\caption{Classification of dynamical regimes. The data possess noise with an amplitude of $0.3$ of the standard deviation ($x_i = x_i + \xi_i$ where $|\xi_i| \le 0.3\sigma(x)$), and $10\%$ of the data is randomly removed. (A) The ground truth classification was computed by the Lyapunov exponent $\Lambda$. The classifications by (B) TACTS and (C) interpolation indicate periodic regions with blue areas and chaotic regions with white areas in the trapezoids.}
\label{fig:classify}
\end{center}
\end{figure}

We assess the fidelity of our method using the following binary classification task: we can compute the maximum Lyapunov exponent ($\Lambda$) of this system for the parameter $r_n$ at each time step, and we classify the system at every point as chaotic if $\Lambda_n>0$ or periodic if $\Lambda_n<0$ (Fig.~\ref{fig:classify}(A)). We select a range of TACTS segment sizes $\{\omega_1,..,\omega_k\}$ and generate the timeline for the vector of costs $c(t)=(c_1,..,c_k)(t)\in \mathbb{R}^k_+$. The set of $\omega_i$ corresponds to segments that contain between 3 and 12 data points. We then chose a set of sliding recurrence frames of lengths $L\in[50,1000]$ and built the spectrum determinism series $SDET$. In order to compare the TACTS method with the interpolation, we first calculated the determinism, DET, series from $\{\hat{x_i}\}$ obtained from using linear interpolation on $\{x_i\}$. We used the following classifier method for both TACTS and interpolated series: If the determinism value is greater than the mean determinism of the series, we classify dynamics around this time as periodic. Conversely, we classify chaotic regimes if a determinism score is less than the mean determinism. The results of this classification method for both series and the ground truth can be seen in Fig.~\ref{fig:classify}\\

\begin{figure}[h!]
\begin{center}
\includegraphics[width=0.88\columnwidth]{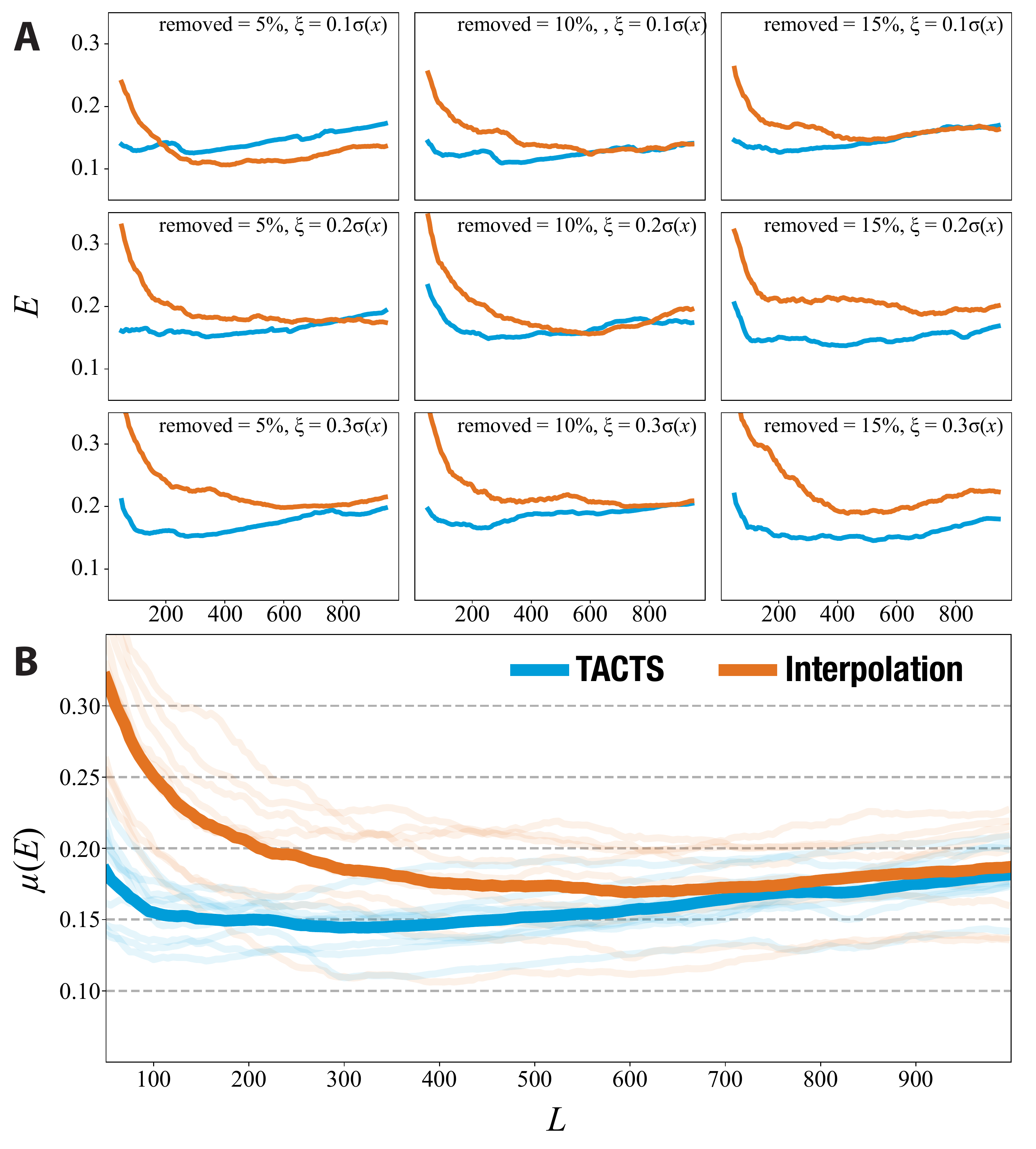}
\captionsetup{width=0.9\textwidth}
\caption{Comparison of classification errors ($E$) between TACTS and interpolation regarding the recurrence frame's length ($L$). Orange lines denote error rates for interpolation, and Blue lines denote error rates for the TACTS series. (A) Logistic map experiments with varying removal ratios and noise levels. (B) The classification error rates averaged across all experiments, $\mu(E)$, for TACTS and interpolation. Transparent traces belong to individual results in (A).}
\label{fig:classify2}
\end{center}
\end{figure}

The mismatch ratios ($E$) between the ground truth (Fig.~\ref{fig:classify}(A)) and regions classified as chaotic or periodic by TACTS (Fig.~\ref{fig:classify}(B)) or interpolated series (Fig.~\ref{fig:classify}(C)) for different levels of data distortion is summarized in Fig.~\ref{fig:classify2}. For all of our test cases except the least distorted example, TACTS provides a classification closer to the Lyapunov exponent classifier for every selected $L$. Even in the case of least data distortion  (Fig.~\ref{fig:classify2}(A)-top left subplot) where TACTS and interpolation have similar classification abilities, the fact that TACTS can better classify the dynamics using a smaller recurrence frame means that it can practically detect more abrupt regime changes. This general trend is observed for the average errors as well. (Fig.~\ref{fig:classify2}(B)).\\

\begin{figure}[h!]
\begin{center}
\includegraphics[width=.83\columnwidth]{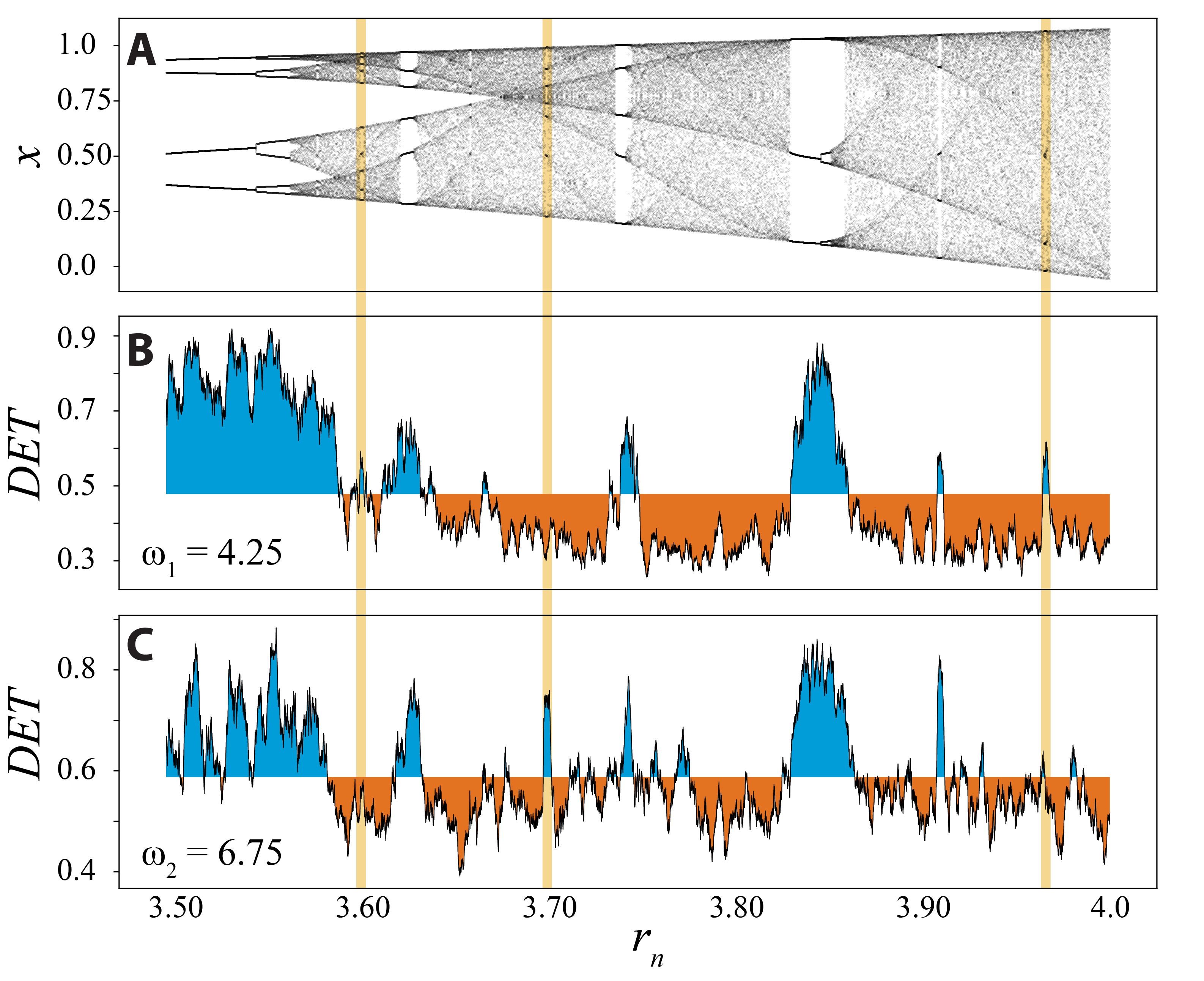}
\captionsetup{width=0.9\textwidth}
\caption{Illustration of the effect of TACTS' segment size $\omega$ with a recurrence window L = 150. Blue (orange) areas indicate that when $DET$ values are higher (lower) than the average $DET$. Yellow vertical lines show different periodic windows for the transient logistic map only detected by specific $\omega$ values. (A) Bifurcation diagram of the logistic map to visualize periodic and chaotic windows. (B) Determinism with $\omega_1=4.25$ successfully detects the period-4 related dynamics around $r\sim3.96$. (C) Determinism from TACTS with $\omega_2=6.75$ identifies the period-7 region around $r\sim3.7$. Note that period-10 region at $r\sim3.6$ is also detected using $\omega_1=4.25$(B) which corresponds to approximately half of the period in this region. }
\label{fig:sensitivity}
\end{center}
\end{figure}

An optimal recurrence frame for each method minimizes the classification error; however, TACTS performs better for a wider range of recurrence frames. This is important since a choice of smaller recurrence frames allows detecting shorter periods of distinct dynamical properties. Note that for this example, it is impossible to find the shortest periodic regions visible in the ground truth obtained from Lyapunov exponent, even with undistorted data since its resolution is low and periodic behavior does not manifest due to existence of transient dynamics. Aggregated determinism series generally provide a more accurate picture but individual TACTS series' determinism also contains added information. While more stable regimes are registered in all determinism series, shorter or more transient regimes are detected only by the TACTS series with segment sizes $\omega$ close to the periodicity of data. Therefore, while we identify periodic dynamical windows, we are also able to estimate the period of these windows. For instance, We can see the determinism series for two TACTS windows generated from the distorted logistic map with $0.1$ standard noise and $10\%$ removal in Fig.~\ref{fig:sensitivity}. Here, we highlighted some regions of dynamics which are reflected differently by different TACTS embeddings. \\

\begin{figure}[h!]
\begin{center}
\includegraphics[width=0.9\columnwidth]{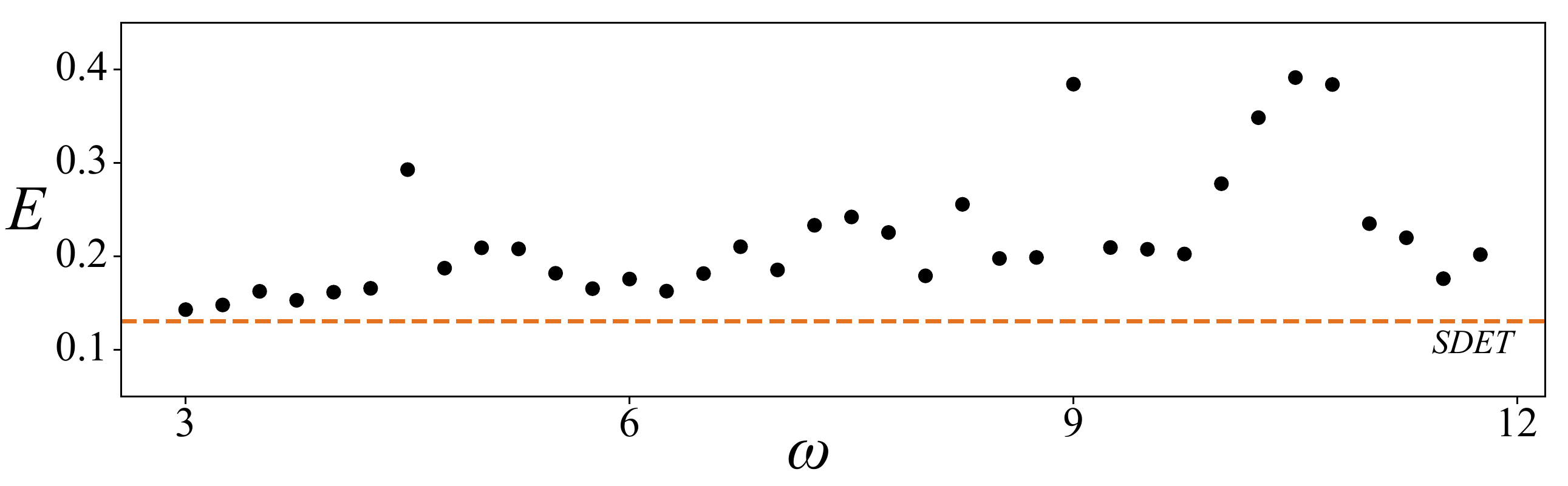}
\captionsetup{width=0.9\textwidth}
\caption{The TACTS' segment size $\omega$  versus the classification error ratios ($E$) associated with the Lyapunov exponent. SDET (orange dashed line) provides a better classification than each individual DET series (black dots). We used the transient logistic map with $0.1$ standard noise and $10\%$ removal for this test with the recurrence window $L=150$.}
\label{fig:advantage}
\end{center}
\end{figure}

We can see that the determinism series obtained from a TACTS series with window $\omega_1=4.25$ in Fig.~\ref{fig:sensitivity}(B) shows peaks in regions of the logistic map with the period-4 around $r=3.96$ and the period-10 region at $r=3.6$. These regions are not identified for $\omega_2=6.75$ (Fig.~\ref{fig:sensitivity}(C)). On the other hand, $DET$ values for $\omega_2=6.75$ allow the identification of the period-7 region at $r=3.7$, which has a period close to its TACTS segment size -- concluding that each TACTS series in the spectrum efficiently captures short transient dynamics with specific periods regarding segment size $\omega$. Nevertheless, using aggregated determinism of the TACTS' spectrum provides a better overall classification performance. Error rates of individual DET classifiers over all points can be seen in Fig. \ref{fig:advantage} compared to error ratio of spectrum determinism $SDET$.\\

\subsection{Real world application}

\begin{figure}[h!]

\includegraphics[width=0.9\columnwidth]{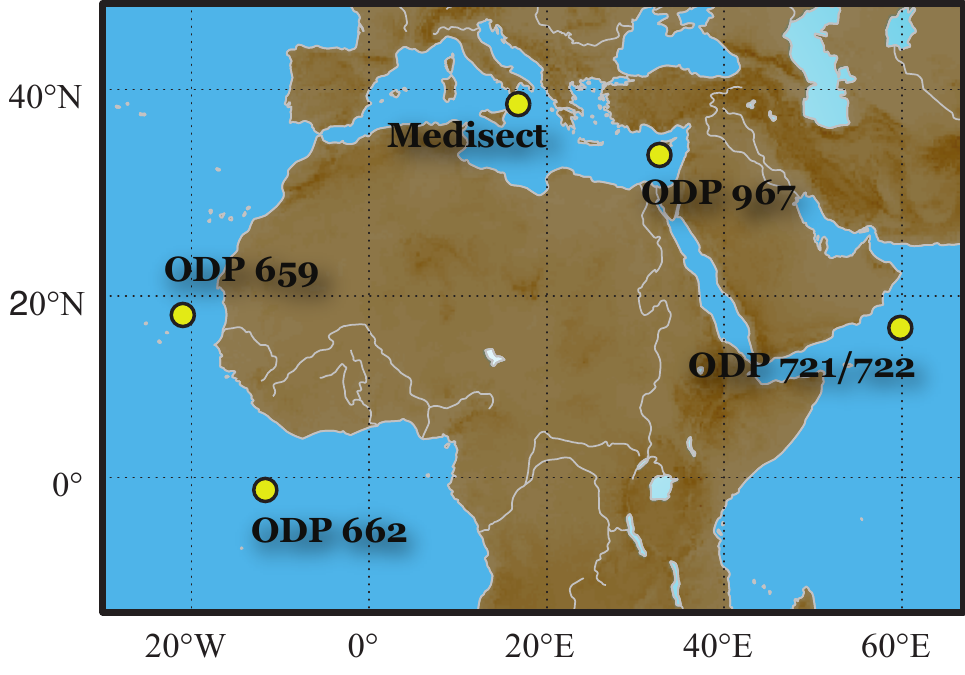}
\captionsetup{width=0.9\textwidth}
\caption{Locations of the proxy archives used in this study.}
\label{fig:map}

\end{figure}

An important application of the TACTS method is the identification of regime changes in palaeoclimate data \cite{eroglu2016,marwan2018tsonis}. Palaeoclimate archives yield irregular time series since the archives accumulate with a time-dependent sedimentation rate and we assume temporal noise since the sampling time points are measurements themselves with potentially large and time-dependent uncertainties. We use several proxy data series obtained from marine sediments located around northern Africa used by previous studies  \cite{demenocal1995plio,trauth2009trends,marwan2021}. Locations of these archives are illustrated in Fig.~\ref{fig:map}. \\

\begin{figure}[h!]

\includegraphics[width=1.03\columnwidth]{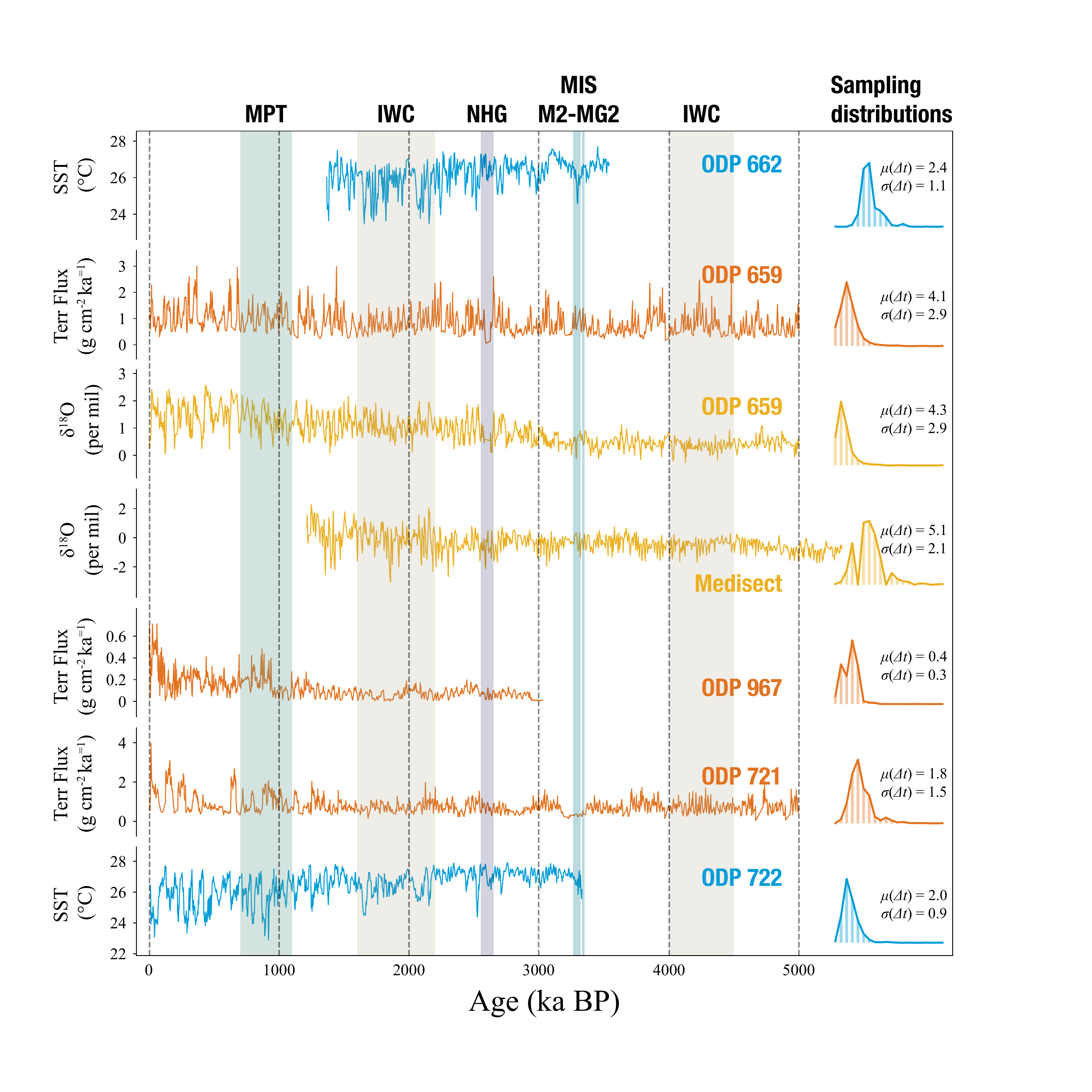}
\begin{center}
\captionsetup{width=0.9\textwidth}
\caption{Data used in this study.  Palaeoclimate time series: blue – sea surface temperature proxies, orange – terrigenous dust flux proxies, yellow - oxygen isotope proxies and important palaeoclimate regimes: Mid-Pleistocene transition (MPT), Northern hemisphere glaciation (NHG), Marine Isotope stage M2 and two distinct periods of intensified Walker circulation (IWC) (left column). Time sampling distributions of time series and their average time step $\mu(\Delta t)$ and their standard deviations $\sigma(\Delta t)$ (right column).}
\label{fig:data}
\end{center}
\end{figure}

We consider 7 data series coming from 3 different types of proxies: Terrigenous dust flux (Dust) for the aridification, alkenone-based sea surface temperature (SST) for the regional temperature, and benthic $\delta^{18}O$ for global ice volume. These data sets and the period they span are illustrated on a single timeline in Fig.~\ref{fig:data} next to their effective sampling distributions. \\

The timeline in Fig.~\ref{fig:data} contains several established climate regime transitions. The transition from Pliocene to the Pleistocene 2.6 million years ago was signified by the onset of periodic Northern Hemisphere Glaciation (NHG). This transition follows a significant climate reorganization during the Pliocene with the establishment of a strong Walker circulation denoted by IWC (Intensified Walker Circulation) between 4500 and 4000 ka BP \cite{ravelo2004regional} and the period of anomalously cold marine isotope stage M2 at 3300 ka BP \cite{tiedemann1994astronomic,lisiecki2005pliocene} which contains the first occurrence of partial glaciation on the northern hemisphere (MG2) regarded as a failed attempt by the climate system to establish an ice age \cite{shackleton199515}. Then another regime, accompanying strong and shifted atmospheric Walker circulation between 2200 and 1500 ka BP \cite{ravelo2004regional} occurred. Finally, Mid-Pleistocene Transition (MPT) took place between 1100 and 700 ka BP, which is characterized by the change in the periodicity of the glacial cycles from approximately 41 ka to a 100 ka dominant periodicity \cite{clark2006middle}.\\

Data properties and the parameters used in the analysis for both TACTS and RP stages are located in Table~1, which also includes our data sources' references. We evaluate all TACTS series on the regular timeline with $\Delta t=1$ ka. For each proxy data series, we choose a number ($N_\omega$) of TACTS window sizes $\omega_i$ using data characteristics and sampling envelopes for each proxy. The shortest $\omega$-windows are chosen such that their corresponding TACTS series encode only a few empty data segments, and these gaps do not appear consecutively. The longest $\omega$-windows are chosen to allow for enough independent values for the corresponding TACTS series. In all the proxy series except one, this procedure eliminated most of the viable spectrum due to a small number of long gaps in the data. In these cases, we sacrificed a portion of our overlapping recurrence timeline containing such problematic regions and repeated the procedure described above, disregarding the points in these regions. Table~1 lists the number of such gaps that had to be created in the recurrence timeline as well as the minimum and maximum period lengths associated with the choice of $\omega_i$ for each proxy data. All determinism series are evaluated with a recurrence frame size $L = 200$ ka and along the shared timeline $T_\rho=\{0, 5, 10, ..., 5000\}$(ka). The number of determinism values calculated from the TACTS series of proxy data evaluated along this timeline is denoted by $N_\rho$. This number is reduced to $N_{\rho*}$ in the presence of gaps, as we accommodate several distinct partial recurrence timelines. \\

\begin{center}
\begin{table}[]
\label{tab:parameters}
\captionsetup{width=0.9\columnwidth}
\caption{Characteristics of the proxy archive data set, the parameters used for their analyses and the original data references.}
\begin{center}
\resizebox{0.9\textwidth}{!}{
\begin{tabular}{llllllll}
\textbf{}                          & \textbf{ODP662}                & \textbf{ODP659T}               & \textbf{ODP659I}               & \textbf{Medisect}              & \textbf{ODP967}                & \textbf{ODP721}                & \textbf{ODP722}                \\ \hline
\multicolumn{1}{|l|}{Type}         & \multicolumn{1}{l|}{SST}       & \multicolumn{1}{l|}{Dust}      & \multicolumn{1}{l|}{Iso}       & \multicolumn{1}{l|}{Iso}       & \multicolumn{1}{l|}{Dust}      & \multicolumn{1}{l|}{Dust}      & \multicolumn{1}{l|}{SST}       \\ \hline
\multicolumn{1}{|l|}{$T$ (Ma)}     & \multicolumn{1}{l|}{3.54-1.37} & \multicolumn{1}{l|}{5.00-0}    & \multicolumn{1}{l|}{5.00-0}    & \multicolumn{1}{l|}{5.33-1.21} & \multicolumn{1}{l|}{3.03-0}    & \multicolumn{1}{l|}{5.00-0}    & \multicolumn{1}{l|}{3.33-0}    \\ \hline
\multicolumn{1}{|l|}{$N$}            & \multicolumn{1}{l|}{360}       & \multicolumn{1}{l|}{1221}      & \multicolumn{1}{l|}{1170}      & \multicolumn{1}{l|}{811}       & \multicolumn{1}{l|}{8417}      & \multicolumn{1}{l|}{2757}      & \multicolumn{1}{l|}{1680}      \\ \hline
\multicolumn{1}{|l|}{$\mu(\Delta t)$ (ka)}    & \multicolumn{1}{l|}{2.39}      & \multicolumn{1}{l|}{4.10}      & \multicolumn{1}{l|}{4.28}      & \multicolumn{1}{l|}{5.08}      & \multicolumn{1}{l|}{0.36}      & \multicolumn{1}{l|}{1.81}      & \multicolumn{1}{l|}{1.98}      \\ \hline
\multicolumn{1}{|l|}{$\sigma(\Delta t)$ (ka)} & \multicolumn{1}{l|}{1.05}      & \multicolumn{1}{l|}{2.69}      & \multicolumn{1}{l|}{2.88}      & \multicolumn{1}{l|}{2.06}      & \multicolumn{1}{l|}{0.31}      & \multicolumn{1}{l|}{1.52}      & \multicolumn{1}{l|}{0.89}      \\ \hline
                                   &                                &                                &                                &                                &                                &                                &                                \\
\textbf{TACTS}                     &                                &                                &                                &                                &                                &                                &                                \\ \hline
\multicolumn{1}{|l|}{minP (ka)}     & \multicolumn{1}{l|}{7.16}      & \multicolumn{1}{l|}{16.38}     & \multicolumn{1}{l|}{12.83}     & \multicolumn{1}{l|}{15.25}     & \multicolumn{1}{l|}{5.40}      & \multicolumn{1}{l|}{7.25}      & \multicolumn{1}{l|}{5.94}      \\ \hline
\multicolumn{1}{|l|}{maxP (ka)}     & \multicolumn{1}{l|}{22.67}     & \multicolumn{1}{l|}{38.91}     & \multicolumn{1}{l|}{40.62}     & \multicolumn{1}{l|}{48.30}     & \multicolumn{1}{l|}{10.61}     & \multicolumn{1}{l|}{17.21}     & \multicolumn{1}{l|}{18.80}     \\ \hline
\multicolumn{1}{|l|}{$\omega$}        & \multicolumn{1}{l|}{3-9.5}     & \multicolumn{1}{l|}{4-9.5}     & \multicolumn{1}{l|}{3-9.5}     & \multicolumn{1}{l|}{3-9.5}     & \multicolumn{1}{l|}{15-29.5}   & \multicolumn{1}{l|}{4-9.5}     & \multicolumn{1}{l|}{3-9.5}     \\ \hline
\multicolumn{1}{|l|}{$N_\omega$}        & \multicolumn{1}{l|}{14}     & \multicolumn{1}{l|}{12}     & \multicolumn{1}{l|}{14}     & \multicolumn{1}{l|}{14}     & \multicolumn{1}{l|}{30}   & \multicolumn{1}{l|}{12}     & \multicolumn{1}{l|}{14}     \\ \hline

\multicolumn{1}{|l|}{Gaps}         & \multicolumn{1}{l|}{1}         & \multicolumn{1}{l|}{2}         & \multicolumn{1}{l|}{1}         & \multicolumn{1}{l|}{0}         & \multicolumn{1}{l|}{2}         & \multicolumn{1}{l|}{3}         & \multicolumn{1}{l|}{1}         \\ \hline
                                   &                                &                                &                                &                                &                                &                                &                                \\
\textbf{RP}                        &                                &                                &                                &                                &                                &                                &                                \\ \hline
\multicolumn{1}{|l|}{$T_\rho$(Ma)}  & \multicolumn{1}{l|}{3.35-1.55} & \multicolumn{1}{l|}{4.80-0.20} & \multicolumn{1}{l|}{4.80-0.20} & \multicolumn{1}{l|}{5.15-1.40} & \multicolumn{1}{l|}{2.85-0.20} & \multicolumn{1}{l|}{4.80-0.40} & \multicolumn{1}{l|}{3.15-0.20} \\ \hline
\multicolumn{1}{|l|}{$N_\rho$}           & \multicolumn{1}{l|}{360}       & \multicolumn{1}{l|}{920}       & \multicolumn{1}{l|}{920}       & \multicolumn{1}{l|}{750}       & \multicolumn{1}{l|}{530}       & \multicolumn{1}{l|}{880}       & \multicolumn{1}{l|}{590}       \\ \hline
\multicolumn{1}{|l|}{$N_{\rho*}$}          & \multicolumn{1}{l|}{349}       & \multicolumn{1}{l|}{787}       & \multicolumn{1}{l|}{851}       & \multicolumn{1}{l|}{750}       & \multicolumn{1}{l|}{406}       & \multicolumn{1}{l|}{697}       & \multicolumn{1}{l|}{528}       \\ \hline
                                   &                                &                                &                                &                                &                                &                                &                                \\ \hline
\multicolumn{1}{|l|}{\textbf{REF}} & \multicolumn{1}{l|}{\cite{herbert2010tropical}}          & \multicolumn{1}{l|}{\cite{tiedemann1994astronomic}}          & \multicolumn{1}{l|}{\cite{tiedemann1994astronomic}}          & \multicolumn{1}{l|}{\cite{lourens1996evaluation}}          & \multicolumn{1}{l|}{\cite{larrasoana2003three}}          & \multicolumn{1}{l|}{\cite{demenocal1995plio}\cite{peter2004african}}          & \multicolumn{1}{l|}{\cite{demenocal1995plio}\cite{peter2004african}}          \\ \hline
\end{tabular}
}
\end{center}
\end{table} 
\end{center}

To test for statistical significance in the determinism values associated with the proxy data, we use the method of bootstrapping as outlined in Ref.~\cite{marwan2013}. We create surrogate determinism scores bootstrapped to our data properties as follows: (1) we record the probability distribution of the lengths of diagonal structures, $P(\epsilon,l_i)$ in the recurrence plots of the TACTS series at each point in our timeline $T_\rho$ and the number of these structures, including isolated points which correspond to diagonal structures with $l_i=1$. (2) We calculate a number of surrogate determinism values for each TACTS series using the average number of diagonal structures we encountered and with lengths coming from the average probability distribution of diagonal line lengths among all time points, $E(P(\epsilon,l_i))$. (3) Determinism surrogate values calculated from these distributions are then used to build a two sided confidence interval by using 1\% quantiles of the value distribution. (4) Since we aggregate all DET series into a single measure SDET, we also aggregate the confidence intervals for each DET series into an averaged confidence interval. We identify the SDET values outside this interval as significant. \\

\begin{figure}[p]

\includegraphics[width=1.03\columnwidth]{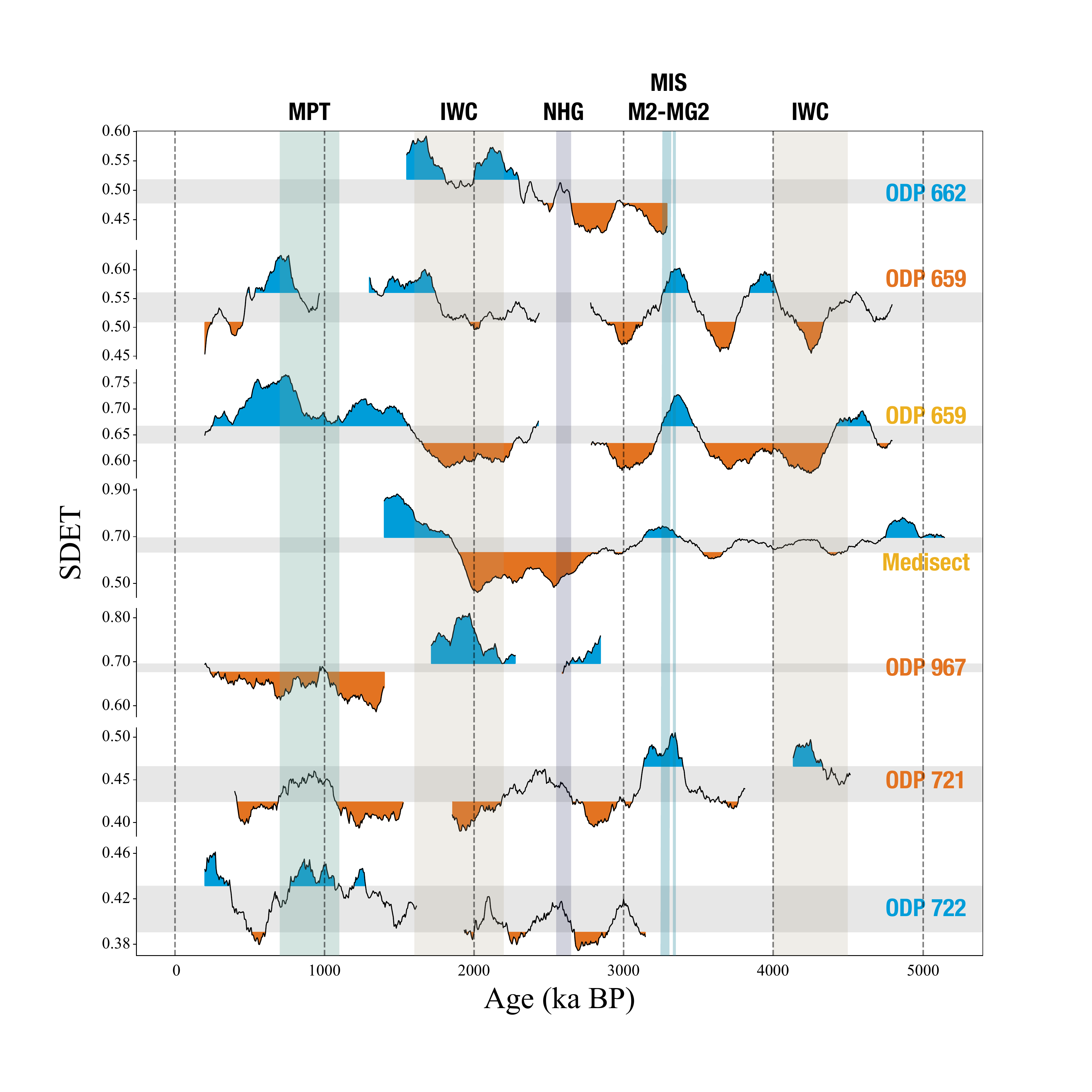}
\captionsetup{width=0.9\columnwidth}
\caption{Spectrum of determinism series generated from the proxy data span a period of about 5 Million years. Each proxy time series is encoded by a number of TACTS series with viable window sizes given in Table 1. and we take the average determinism given by the ensemble as $SDET$. We have visible gaps in some $SDET$ series in periods where the proxy data is too sparse to translate into a meaningful TACTS series. Confidence intervals for the determinism values is displayed with gray horizontal bands which correspond to 1st-99th percentiles. Values of determinism which lie outside this interval are the extremities. We denote regions of high determinism/predictability with the color blue and low-determinism or unpredictable regions with the color orange. }
\label{fig:odpdets}

\end{figure}

Spectrum determinism series generated for our proxies are shown in Fig.\ref{fig:odpdets} where the time is measured in ka BP (thousand years before present) and chronological order is from right to left. In Fig.\ref{fig:odpdets}, we identify several critical periods. Pleistocene IWC starting from 4500 ka BP is characterized by high predictability in dust records from east of Africa (ODP721) while the Atlantic coast proxies show highly chaotic dynamics (ODP659 dust and $\delta^{18}O$). Another strongly accentuated period we detected is the Marine Isotope Stage M2 around 3300 ka BP where all of $\delta^{18}O$ and Dust proxies from both sides of the continent (ODP659,ODP721) and Mediterranean (Medisect) report very predictable behavior. In particular, we found a dominant $\omega$-window corresponding to a 10 ka cycle driving the Arabian and Mediterranean seas during the M2 cooling event.\\

Prior to 2600 ka BP (NHG period), Eastern Africa (ODP721/722) proxies start to display significantly low predictability followed by a significant drop of predictability in the Mediterranean (Medisect, ODP967) during NHG; however, the west side is yet unaffected (ODP659). This transition over the Mediterranean signifies a re-organization of regional climate actors until 2200 ka BP, after which Atlantic coast dust records (ODP659) show a long period of unpredictability, coinciding with the strong Walker circulation period (IWC) driving the ocean and stabilizing temperature cycles (ODP662). The transition period after NHG is followed by the consequent stabilization of the Mediterranean climate during the IWC event, starting from the east Mediterranean. (ODP967, Medisect)\\

During the Mid-Pleistocene Transition (MPT), starting at 1100 ka BP, we see a gradual increase in the determinism of Dust and SST proxies located in the eastern parts of Africa (ODP 721/722) and the eastern Mediterranean (ODP967). This is mirrored by a corresponding gradual decrease of determinism on the west coast (ODP659 $\delta^{18}O$). While individual determinism values do not clearly indicate a regime transition (ODP659 determinism stays significantly high and ODP967 determinism stays significantly low), persistent anti-correlation between these clusters during this period suggests a transient see-saw dynamics between east and west. After the MPT, this strong anti-correlation breaks, and Arabian sea proxies (ODP721/722) start to signal unpredictable dynamics again.\\

The same proxy data set is analyzed using interpolation with several analysis methods, including windowed determinism $DET(t)$ analysis, Ref.~\cite{marwan2021}. Although there is a general agreement between our results and $DET$ results of Ref.~\cite{marwan2021}, there are also some crucial differences. We found the Arabian Sea (ODP721) to be relatively predictable during the Pliocene IWC, while it is significantly unpredictable and unresponsive to IWC in Ref.~\cite{marwan2021}. Both results agree with the indication of the cooling event M2; however, we also found a more pronounced transition to order in the Mediterranean (Medisect) accompanying partial glaciation. While our findings agree with Ref.\cite{marwan2021} on the loss of predictability in Africa after the onset of NHG,  our analysis suggests a longer reign of low determinism in the Mediterranean (which we find to be highly deterministic at other times) until the transition occurs. In contrast, analysis of Ref.~\cite{marwan2021} shows high predictability in this period compared to a chaotic pre-NHG Mediterranean. Our application employed the TACTS spectrum and used a smaller recurrence frame (200 ka instead of 410 ka). Thus, we observed smoother trends in determinism series, specifically during Mid-Pleistocene Transition and IWC events.

\section{Conclusion}
\label{sec:conclusion}
This paper introduces modifications for the TACTS method for analyzing irregularly sampled time series by transforming the data into regularly sampled cost time series. We have evaluated the modified TACTS method's performance on the logistic map and analyzed palaeoclimate records surrounding Africa to demonstrate the method's usefulness.\\

The fundamental modification is to construct a spectrum of TACTS time series following elementary three point-wise operations: (1) scaling the amplitude, (2) shifting a data point in time, and (3) ignoring a data point. Each operation incurs a cost with selected factors. While the cost coefficients of operations (1) and (2) directly depend on the associated time series and the TACTS timeline, the cost of operation (3) is a free parameter to optimize our metric. Note that this free parameter allows us to tweak a suitable and information-preserving time series preprocessing method for further analysis without the corruption risk of interpolation methods. The total cost of transforming neighboring data segments to each other determines the TACTS point; applying this procedure over the entire data generates TACTS for an arbitrarily selected segment size ($\omega$). However, different segment sizes $\omega$ lead to changes in the resulting cost time series. We demonstrated that $\omega$ value behaves as an embedding dimension \cite{packard80}, and varying $\omega$ allows detecting different periodic regions in the system's phase space. Therefore, we introduced analyzing the TACTS spectrum of various $\omega$ values to detect all regime transitions.\\

The performance of the modified TACTS has demonstrated that the approach is functional even for extremely irregular sampling and high measurement noise. Blending the spectrum of TACTS with RP analysis, we accurately detected dynamical regime changes in the logistic map whose control parameter systematically changes in time. The systematic comparison of the performances of interpolation and TACTS in this test showed that TACTS outperforms the interpolation. \\

Applying the TACTS and RP approach to palaeoclimate data surrounding Africa, we identified various important regime changes in the climate dynamics during the last 5,000,000 years. We provided a timeline consistent with this period's well-documented transitions. Detailed climatological interpretation of our findings is the subject of an ongoing study collaborating with climatologists and will be published in more appropriate climate journals.\\

An irregular sampling of proxy records is a natural fact in Earth science where the temporal data comes from the measurement process of aging and uneven growth rates of proxy structures and sedimentation rates result in irregularly spaced time points. Furthermore, the proxy records may have poor quality and resolution due to unavoidable damages over the lifetime of the historical structures and scanning expenses. Therefore, the TACTS technique has significant potential in quantitative Earth science to gently preprocess the data and prepare it for analysis by modern time series analysis techniques. 

\section*{Acknowledgments}
We thank Norbert Marwan and Sebastian F.~M.~Breitenbach for enlightening discussions. Financial support from TUBITAK is acknowledged (Grant No. 118C236). D.E. was supported by the BAGEP Award of the Science Academy. 

\section*{Data Availability Statement}
This manuscript has associated data in a data repository. TACTS code library for the analysis can be accessed via \url{https://doi.org/10.5281/zenodo.6038896}. The proxies are published/available data sets. Requests for the TACTS of proxies can be sent to \href{mailto:celik.ozdes@khas.edu.tr}{celik.ozdes@khas.edu.tr}.

\bibliographystyle{epj}
\bibliography{TACTS}

\begin{thebibliography}{45}

\bibitem{anderson1977box}
O.~Anderson, RAIRO-Operations Research \textbf{11}, 3 (1977)

\bibitem{friedman2001}
J.~Friedman, T.~Hastie, R.~Tibshirani et~al., \emph{The elements of statistical
  learning}, Vol.~1 (Springer series in statistics New York, 2001)

\bibitem{livina2007}
V.N. Livina, T.M. Lenton, Geophysical Research Letters \textbf{34}, L03712
  (2007)

\bibitem{trulla96}
L.L. Trulla, A.~Giuliani, J.P. Zbilut, C.L. {Webber Jr.}, Physics Letters A
  \textbf{223}, 255 (1996)

\bibitem{small2013complex}
M.~Small, \emph{Complex networks from time series: Capturing dynamics}, in
  \emph{2013 IEEE International Symposium on Circuits and Systems (ISCAS)}
  (IEEE, 2013), pp. 2509--2512

\bibitem{marwan2021}
N.~Marwan, J.F. Donges, R.V. Donner, D.~Eroglu, Quaternary Science Reviews
  \textbf{274}, 107245 (2021)

\bibitem{eroglu2014a}
D.~Eroglu, N.~Marwan, S.~Prasad, J.~Kurths, Nonlinear Processes in Geophysics
  \textbf{21}, 1085 (2014)

\bibitem{rehfeld2013}
K.~Rehfeld, N.~Marwan, S.F.M. Breitenbach, J.~Kurths, Climate Dynamics
  \textbf{41}, 3 (2013)

\bibitem{eroglu2016}
D.~Eroglu, F.H. McRobie, I.~Ozken, T.~Stemler, K.H. Wyrwoll, S.F. Breitenbach,
  N.~Marwan, J.~Kurths, Nature communications \textbf{7}, 1 (2016)

\bibitem{rehfeld2011}
K.~Rehfeld, N.~Marwan, J.~Heitzig, J.~Kurths, Nonlinear Processes in Geophysics
  \textbf{18}, 389 (2011)

\bibitem{rehfeld2014}
K.~Rehfeld, J.~Kurths, Climate of the Past \textbf{10}, 107 (2014)

\bibitem{ozken2015}
I.~Ozken, D.~Eroglu, T.~Stemler, N.~Marwan, G.B. Bagci, J.~Kurths, Physical
  Review E \textbf{91}, 062911 (2015)

\bibitem{ozken2018}
I.~Ozken, D.~Eroglu, S.F.M. Breitenbach, N.~Marwan, L.~Tan, U.~Tirnakli,
  J.~Kurths, Physical Review E \textbf{98}, 052215 (2018)

\bibitem{marwan2007}
N.~Marwan, M.C. Romano, M.~Thiel, J.~Kurths, Physics Reports \textbf{438}, 237
  (2007)

\bibitem{victor1997}
J.D. Victor, K.P. Purpura, Network: Computation in Neural Systems \textbf{8},
  127 (1997)

\bibitem{suzuki2010}
S.~Suzuki, Y.~Hirata, K.~Aihara, International Journal of Bifurcation and Chaos
  \textbf{20}, 3699 (2010)

\bibitem{braun2021sampling}
T.~Braun, C.N. Fernandez, D.~Eroglu, A.~Hartland, S.F. Breitenbach, N.~Marwan,
  arXiv preprint arXiv:2112.04843  (2021)

\bibitem{priestley1996wavelets}
M.~Priestley, Journal of Time Series Analysis \textbf{17}, 85 (1996)

\bibitem{kantz1994robust}
H.~Kantz, Physics letters A \textbf{185}, 77 (1994)

\bibitem{pincus1991approximate}
S.M. Pincus, Proceedings of the National Academy of Sciences \textbf{88}, 2297
  (1991)

\bibitem{lacasa2008time}
L.~Lacasa, B.~Luque, F.~Ballesteros, J.~Luque, J.C. Nuno, Proceedings of the
  National Academy of Sciences \textbf{105}, 4972 (2008)

\bibitem{eckmann1987}
J.P. Eckmann, S.~{Oliffson Kamphorst}, D.~Ruelle, Europhysics Letters
  \textbf{4}, 973 (1987)

\bibitem{hirata2010identifying}
Y.~Hirata, K.~Aihara, Physical Review E \textbf{81}, 016203 (2010)

\bibitem{eroglu2014b}
D.~Eroglu, T.K.D. Peron, N.~Marwan, F.A. Rodrigues, L.d.F. Costa, M.~Sebek,
  I.Z. Kiss, J.~Kurths, Phys. Rev. E \textbf{90}, 042919 (2014)

\bibitem{hirata2012timing}
Y.~Hirata, K.~Aihara, Physica A: Statistical Mechanics and its Applications
  \textbf{391}, 760 (2012)

\bibitem{zimatore2021recurrence}
G.~Zimatore, L.~Falcioni, M.C. Gallotta, V.~Bonavolont{\`a}, M.~Campanella,
  M.~De~Spirito, L.~Guidetti, C.~Baldari, PloS one \textbf{16}, e0249504 (2021)

\bibitem{wu2022recurrence}
J.~Wu, X.~Zhou, Y.~Peng, X.~Zhao, Physica A: Statistical Mechanics and its
  Applications \textbf{585}, 126439 (2022)

\bibitem{proksch2022coordination}
S.~Proksch, M.~Reeves, M.~Spivey, R.~Balasubramaniam, Scientific reports
  \textbf{12}, 1 (2022)

\bibitem{baranowski2022being}
G.~Baranowski-Pinto, V.~Profeta, M.~Newson, H.~Whitehouse, D.~Xygalatas,
  Scientific reports \textbf{12}, 1 (2022)

\bibitem{poincare1890}
H.~Poincar\'e, Acta Mathematica \textbf{13}, 1 (1890)

\bibitem{addo2013}
P.M. Addo, M.~Billio, D.~Guegan, The North American Journal of Economics and
  Finance \textbf{26}, 416 (2013)

\bibitem{marwan2018tsonis}
N.~Marwan, D.~Eroglu, I.~Ozken, T.~Stemler, K.H. Wyrwoll, J.~Kurths, in
  \emph{Advances in Nonlinear Geosciences}, edited by A.A. Tsonis (Springer
  International, Cham, Switzerland, 2018), pp. 357--368, ISBN 978-3-319-58895-7

\bibitem{demenocal1995plio}
P.B. Demenocal, Science \textbf{270}, 53 (1995)

\bibitem{trauth2009trends}
M.H. Trauth, J.C. Larrasoana, M.~Mudelsee, Quaternary Science Reviews
  \textbf{28}, 399 (2009)

\bibitem{ravelo2004regional}
A.C. Ravelo, D.H. Andreasen, M.~Lyle, A.O. Lyle, M.W. Wara, Nature
  \textbf{429}, 263 (2004)

\bibitem{tiedemann1994astronomic}
R.~Tiedemann, M.~Sarnthein, N.J. Shackleton, Paleoceanography \textbf{9}, 619
  (1994)

\bibitem{lisiecki2005pliocene}
L.E. Lisiecki, M.E. Raymo, Paleoceanography \textbf{20} (2005)

\bibitem{shackleton199515}
N.~Shackleton, M.~Hall, D.~Pate, \emph{15. Pliocene stable isotope stratigraphy
  of Site 846}, in \emph{Proc. Ocean Drill. Program Sci. Results} (1995), Vol.
  138, pp. 337--355

\bibitem{clark2006middle}
P.U. Clark, D.~Archer, D.~Pollard, J.D. Blum, J.A. Rial, V.~Brovkin, A.C. Mix,
  N.G. Pisias, M.~Roy, Quaternary Science Reviews \textbf{25}, 3150 (2006)

\bibitem{herbert2010tropical}
T.D. Herbert, L.C. Peterson, K.T. Lawrence, Z.~Liu, science \textbf{328}, 1530
  (2010)

\bibitem{lourens1996evaluation}
L.J. Lourens, A.~Antonarakou, F.~Hilgen, A.~Van~Hoof, C.~Vergnaud-Grazzini,
  W.~Zachariasse, Paleoceanography \textbf{11}, 391 (1996)

\bibitem{larrasoana2003three}
J.~Larrasoa{\~n}a, A.~Roberts, E.~Rohling, M.~Winklhofer, R.~Wehausen, Climate
  Dynamics \textbf{21}, 689 (2003)

\bibitem{peter2004african}
B.d. Peter, Earth and Planetary Science Letters \textbf{220}, 3 (2004)

\bibitem{marwan2013}
N.~Marwan, S.~Schinkel, J.~Kurths, Europhysics Letters \textbf{101}, 20007
  (2013)

\bibitem{packard80}
N.H. Packard, J.P. Crutchfield, J.D. Farmer, R.S. Shaw, Physical Review Letters
  \textbf{45}, 712 (1980)

\end{thebibliography}

\end{document}